\documentclass[
reprint,
superscriptaddress,
showpacs,
preprintnumbers,
nofootinbib,
nobibnotes,
amsmath,
amssymb, 
aps,
prd,
floatfix
]{revtex4-1}

\usepackage[utf8]{inputenc}
\usepackage[normalem]{ulem}
\usepackage{graphicx}
\usepackage{dcolumn}
\usepackage{bm}
\usepackage[colorlinks=true,allcolors=blue]{hyperref}
\usepackage{url}
\usepackage{enumerate}

\usepackage{slashed,multirow,relsize,soul,tikz}
\usepackage{color}
\usepackage{mathrsfs} 

\usepackage{bbold}
 \usepackage{mathrsfs}
\usepackage{braket}
\usepackage{physics}
\usepackage{multirow}

\usepackage{fontawesome} 
\definecolor{blue-violet}{rgb}{0.33, 0.17, 0.89}

\newcommand{\refeq}[1]{Eq.~(\ref{#1})}

\newcommand{\reffig}[1]{Fig.~\ref{#1}}

\newcommand{\refsec}[1]{Section~\ref{#1}}
\newcommand{\refapp}[1]{Appendix~\ref{#1}}
\newcommand{\reftab}[1]{Table~\ref{#1}}
\newcommand{\refref}[1]{Ref.~\cite{#1}}

\def\eg{\emph{e.g.}}

\newcounter{CommentCount}
\definecolor{MH}{rgb}{0.0,0.6,9}
\definecolor{palatinate}{rgb}{0.494, 0.192, 0.482}

\renewcommand{\phi}{\varphi}

\begin{document}

\preprint{\hfill FTPI-MINN-20-28}

\title{Constraints on Decaying Sterile Neutrinos from Solar Antineutrinos}

\author{Matheus Hostert}
\email{mhostert@umn.edu}
\affiliation{School of Physics and Astronomy, University of Minnesota, Minneapolis, MN 55455, USA}
\affiliation{William I. Fine Theoretical Physics Institute, School of Physics and Astronomy, University of
Minnesota, Minneapolis, MN 55455, USA}
\affiliation{Perimeter Institute for Theoretical Physics, Waterloo, ON N2J 2W9, Canada}

\author{Maxim Pospelov}
\affiliation{School of Physics and Astronomy, University of Minnesota, Minneapolis, MN 55455, USA}
\affiliation{William I. Fine Theoretical Physics Institute, School of Physics and Astronomy, University of
Minnesota, Minneapolis, MN 55455, USA}

\date{\today}

\begin{abstract}
  Solar neutrino experiments are highly sensitive to sources of $\nu\to\overline{\nu}$ conversions in the $^8$B neutrino flux. In this work we adapt these searches to non-minimal sterile neutrino models recently proposed to explain the LSND, MiniBooNE, and reactor anomalies. The production of such sterile neutrinos in the Sun, followed the decay chain $\nu_4 \to \nu \phi \to \nu \nu \overline{\nu}$ with a new scalar $\phi$ results in upper limits for the neutrino mixing $|U_{e4}|^2$ at the per mille level. We conclude that a simultaneous explanations of all anomalies is in tension with KamLAND, Super-Kamiokande, and Borexino constraints on the flux of solar antineutrinos. We then present other minimal models that violate parity or lepton number, and discuss the applicability of our constraints in each case. Future improvements can be expected from existing Borexino data as well as from future searches at Super-Kamiokande with added Gd.
\end{abstract}

\maketitle

\section{Introduction} 
Beyond neutrino mixing, the study of solar neutrinos provides important input to the standard solar Model (SSM)~\cite{Bahcall:2005va,Agostini:2020mfq} and has been used to search for several phenomena beyond the SM of particle physics. One example is neutrino decay, originally proposed as an alternative solution to the solar neutrino problem~\cite{Bahcall:1972my,Pakvasa:1972gz}. Indeed, after precision measurements of the solar neutrino oscillation parameters by KamLAND~\cite{Abe:2008aa}, strong constraints on the lifetimes of $\nu_2$ and $\nu_3$ have been obtained~\cite{Joshipura:2002fb,Beacom:2002cb} as data is consistent with no additional neutrino disappearance. 

Recently, non-minimal neutrino decay models have received interest in the literature, where exotic decays of a relatively heavy ($m_{\rm exotic} \gg m_{\rm active}$) and mostly-sterile neutrino are invoked to explain longstanding experimental anomalies at short baselines (SBL). One category of models concerns ``visible" sterile neutrino decay, where new sterile states are produced and decay back to visible active neutrinos. Originally proposed in Ref.~\cite{PalomaresRuiz:2005vf} as an explanation to the Liquid Scintillator Neutrino Detector (LSND) anomaly~\cite{Athanassopoulos:1996jb,Aguilar:2001ty}, this scenario has now been revisited~\cite{deGouvea:2019qre,Dentler:2019dhz} in light of recent data of short-baseline $\nu_\mu \to \nu_e$ and $\overline{\nu}_\mu \to \overline{\nu}_e$ appearance at MiniBooNE~\cite{AguilarArevalo:2007it,Aguilar-Arevalo:2018gpe,Aguilar-Arevalo:2020nvw}, as well as $\nu_e$ disappearance at reactors~\cite{Mention:2011rk,Dentler:2017tkw}. Due to the small mixing angles required in this explanation, the effects of attenuation in solar neutrino fluxes is small. Yet, the total number of heavy neutrinos produced is large, and if these states undergo sufficiently distinctive decays or scattering inside a detector, they can be searched for. In this article, we point out that if antineutrinos are produced in the decay of these heavy neutrinos, then they are strongly constrained by existing searches for neutrino-antineutrino transitions in solar neutrino experiments. 

The flux of antineutrinos from the Sun at the MeV energies is negligible~\cite{Malaney:1989hs}, which remains an excellent approximation down to tens of keV in energy~\cite{Vitagliano:2017odj}. Combined with the fact that the detection cross section for $\overline{\nu_e}$ is much larger and easier to measure compared to that of $\nu_e$, this makes solar neutrino experiments sensitive to very small fluxes of antineutrinos~\cite{Akhmedov:1991uk,Barbieri:1991ed,Acker:1992eh}. The current sensitivity reaches fluxes as small as a few times $10^{-5}$ of the $^8$B neutrino flux~\cite{Aharmim:2004uf,Collaboration:2011jza,Agostini:2019yuq,Linyan2018,Super-Kamiokande:2020frs,Abe:2021tkw}. These searches have been discussed in the context of new physics, such as large $\nu\to\overline{\nu}$ oscillations. This Lepton number (LN) violating process is rather small in most theories, being suppressed by $(m_\nu/E)^2$, but can be enhanced due to spin-flavor precession~\cite{Lim:1987tk,Akhmedov:1989df}. The latter arises from the coupling of a large neutrino magnetic moment to the solar magnetic field, which induces $\nu_e \to \overline{\nu_x}$ conversions, followed by flavor transitions into $\overline{\nu_e}$ due to matter effects. Another possibility to generate such LN violating signatures is neutrino decay. For instance, neutrino mass models where LN is a spontaneously broken global symmetry predict the existence of a pseudo-goldstone boson $J$, the majoron~\cite{Chikashige:1980ui,Gelmini:1980re}. In these models, solar antineutrinos may be produced from the decay $\nu_2 \to \overline{\nu_1} J$, which is enhanced in dense matter~\cite{Berezhiani:1987gf}. This possibility of production from neutrino decay is, in fact, quite general and can be realized in any LN violating model with neutrinos that decay sufficiently fast, be they $\nu_2$, $\nu_3$, or the new mostly-sterile state $\nu_4$ discussed here (see Ref.~\cite{Berezhiani:1991vk} for an early discussion in the context of a 17 keV sterile neutrino). 

In this work, we explore a new possibility where lepton number can, in fact, be conserved but the decay of a new light boson leads to a large flux of antineutrinos. We derive limits on the electron flavor mixing with $\nu_4$, working only with the gauge-invariant and parity-conserving model of Ref.~\cite{Dentler:2019dhz}. Focusing solely on Dirac neutrinos, we show that our bounds exclude virtually all of the parameter space preferred that can simultaneously explain LSND and MiniBooNE, as well as the region of interest for reactor anomalies. 
They also disfavor most but not all parameter space suggested as a solution to the MiniBooNE anomaly. For models with Majorana neutrinos, the constraints become even stronger due to $\nu_4 \to \overline{\nu} \phi \to \nu \overline{\nu} \overline{\nu}$ decays.

The paper is organized as follows. In \refsec{sec:model} we review the benchmark model for decaying sterile neutrinos, and in \refsec{sec:solantinus} we discuss generic aspects of solar antineutrino searches. The resulting constraints, future prospects, and alternative search methods are then discussed in \refsec{sec:results}. We dedicate \refsec{sec:alternatives} to a survey of minimal alternative models for decaying steriles, and conclude in \refsec{sec:conclusions}.

\section{Decaying Sterile Neutrino}\label{sec:model}
\begin{figure}[t]
    \centering
    \includegraphics[width=0.49\textwidth]{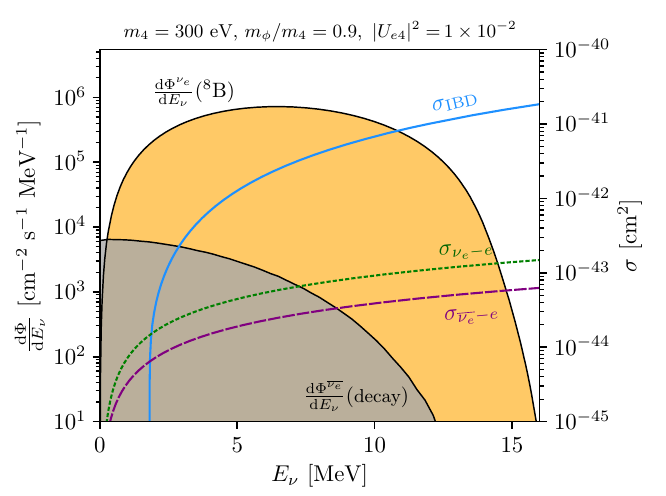}
    \caption{The solar neutrino energy spectrum from $^8$B (shaded orange) and resulting antineutrino spectrum from the decays of $\nu_4$ (shaded grey). We also show the inverse beta decay (IBD) and neutrino-electron scattering cross sections on an overlaid axis. \label{fig:B8flux}}
\end{figure}

The most significant deviations from the three-neutrino paradigm at SBLs are the LSND excess of $\overline{\nu_e}$ events, with a statistical significance of $3.8~\sigma$ when interpreted under a $\overline{\nu}_\mu$ oscillation hypothesis, and the MiniBooNE excess of $\nu_e$-like events, with a significance of $4.8~\sigma$ when interpreted under a ${\nu}_\mu\to {\nu_e}$ and $\overline{\nu}_\mu\to \overline{\nu_e}$ oscillation hypothesis. Reactors at very short-baselines also have some evidence of $\nu_e$ disappearance~\cite{Mention:2011rk,Dentler:2017tkw}, but in that case the neutrino flux predictions are highly uncertain and harder to control~\cite{Berryman:2019hme,Berryman:2020agd}. Despite the large significance of these anomalies, they remain unsolved. Their standard interpretation under oscillations of a eV-scale sterile neutrino leads to strong tensions between different data. This is driven mainly by the absence of anomalous results in $\nu_\mu$ disappearance experiments~\cite{Dentler:2018sju,Diaz:2019fwt}, as appearance and disappearance channels are strongly correlated in the oscillation scenario~\cite{Okada:1996kw,Bilenky:1996rw}. In addition, such new sterile states with eV masses are in strong tension with cosmological observations, which has prompted several studies to resolve this by invoking secret interactions~\cite{Dasgupta:2013zpn,Hannestad:2013ana,Vecchi:2016lty,Farzan:2019yvo,Cline:2019seo}.  

Visible sterile neutrino decays are, therefore, a natural ``next-to-minimal" explanation to SBL anomalies to consider. The advantages of this scenario are that it does not necessarily lead to strong correlations between appearance and disappearance channels, the mass scale of the new sterile state is not fixed by the oscillation length of the experiments, and that it already contains a secret interaction mechanism, possibly alleviating tension with cosmology. In this work, we focus on the decay of steriles with eV to hundreds of keV masses to a new scalar $\phi$, as discussed in Refs.~\cite{PalomaresRuiz:2005vf,Bai:2015ztj,Moss:2017pur,Dentler:2019dhz,deGouvea:2019qre}. In all such visible decay scenarios, heavy neutrinos decay to mostly-active neutrinos via $\nu_4 \to \nu \phi$ , where more neutrinos can be produced from $\phi\to \nu\overline{\nu}$ decay if $\phi$ is massive as in Ref.~\cite{Dentler:2019dhz}. Here \emph{visible} refers to the detectability of the decay products, in contrast to models where neutrinos decay to the wrong-helicity states that do not feel the weak interactions (up to tiny helicity-flipping terms proportional to $m^2_\nu/E_\nu^2$). 

Such visible decays can explain the anomalous $\nu_e$-like events at SBL experiments by means of a sub-dominant population of $\nu_4$ states in neutrino beams, which often decays to $\nu_e$-like daughters~\footnote{Constraints on this scenario have been obtained in Ref.~\cite{Brdar:2020tle} using the near detector of NO$\nu$A and T2K, as well as MINER$\nu$A and PS-191. We note that the constraints have been obtained under simplified assumptions, and that a detailed study with total signal efficiency, as well as appropriate uncertainties is needed in order to derive reliable constraints.}. One typical prediction is that the spectrum of daughter $\nu_e$ and $\overline{\nu_e}$ neutrinos is softer than the initial flux of $\nu_4$ parents and associated neutrinos, skewing the effective flavor conversion towards lower energies. While this brings a mild improvement over the oscillation fit to the low energy excess observed at MiniBooNE, it leads to less satisfactory energy spectra at LSND, which is compatible with a signal that grows in energy. In addition, the neutrino flux at LSND comes from both $\pi^+$ and $\mu^+$ decay at rest, yielding a large and monochromatic $\nu_\mu$ flux, and a spectrum of $\overline{\nu_\mu}$ and $\nu_e$. Since only the $\overline{\nu_e}$ component is detected via the IBD process, the presence of a neutrino-to-antineutrino transition in the decay chain can convert the large $\nu_\mu$ flux to signal, since $\nu_4$ states can be produced in pion as well as muon decays. This is a crucial point in the study of~\refref{Dentler:2019dhz}, which found improved compatibility between LSND and MiniBooNE regions of preference when this conversion is significant.

For concreteness, we focus on the gauge-invariant and parity-conserving model of Ref.~\cite{Dentler:2019dhz}, wherein a SM singlet $\nu_s$ is introduced and equipped with sizable couplings to a new scalar singlet $\phi$. The sterile neutrino can then couple to light and mostly-active neutrinos in a gauge-invariant fashion by means of mixing between the heaviest neutrino state, $\nu_4$, and the active flavors. The relevant Lagrangian is given by
\begin{equation}\label{eq:Lscalar}
 - \mathscr{L} = g_\phi \overline{{\nu_s}} {\nu_s} \phi + \sum_{\alpha,\beta} m_{\alpha \beta} \overline{{\nu_\alpha}} {\nu_\beta}, 
\end{equation}
where the neutrino mass mechanism is left unspecified and assumed to not play a role in the low-energy phenomenology. In the mass basis, the neutrino mass eigenstates are given by $\nu_i = \sum_\alpha U_{\alpha i}^* \nu_\alpha$, with $\alpha\in \{e,\mu,\tau,s\}$, and $U$ a unitary mixing matrix. Under the assumption of parity conservation in the sterile sector, $U$ is identical to the extended Pontecorvo–Maki–Nakagawa–Sakata (PMNS) matrix, now $4\times4$. We return to this issue in \refsec{sec:alternatives}. In the decays of $\nu_4$ and $\phi$, only the three lightest mass states are produced, and so the it is useful to define the low-energy flavor state $\hat{\nu}_s = \sum_{i=1}^3 U_{\alpha i} \nu_i$. For most processes of interest, however, the non-unitarity corrections introduced by working with $\hat{\nu}_s$ instead of the full flavor states $\nu_s$ is small and appears only at order $|U_{\alpha4}|^4$. Unless stated otherwise, we refer to $\hat{\nu}_s$ as simply $\nu$ from now on, as the mass eigenstates have decohered on their way from Sun. The new scalar does not couple directly to the SM, and loop-induced couplings will ultimately depend on the UV completion of the model and its neutrino mass mechanism (see, for instance, Refs.~\cite{Chikashige:1980ui,Xu:2020qek}).
\begin{figure}[t]
    \centering
    \includegraphics[width=0.49\textwidth]{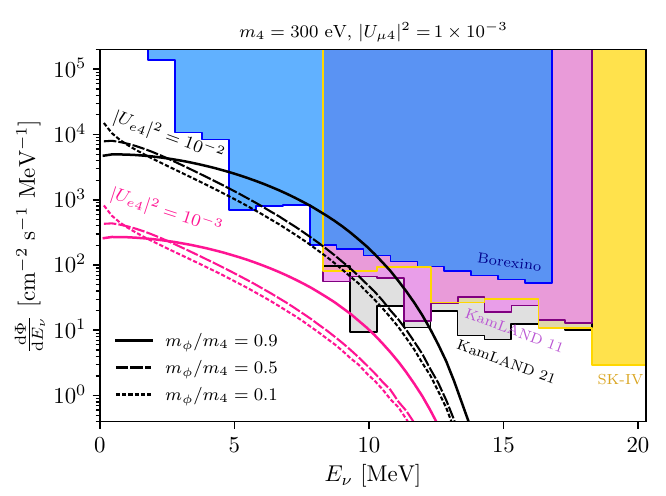}
    \caption{The experimental limits on solar $\overline{\nu_e}$ at $90\%$ C.L. as a function of the neutrino energy. The shaded regions are excluded by Borexino (blue)~\cite{Agostini:2019yuq}, KamLAND 2011~\cite{Collaboration:2011jza} (pink), KamLAND 2021~\cite{Abe:2021tkw} (grey), and SuperK-IV 2020~\cite{Super-Kamiokande:2020frs} (yellow). The different new physics predictions are also shown as solid curves assuming $|U_{\tau 4}|=0$ and $|U_{\mu 4}|^2 = 10^{-3}$. \label{fig:fluxlimits}}
\end{figure}

Due to mixing, heavy neutrinos with masses below the MeV scale would be produced in the Sun via the same processes responsible for $\nu_e$ production at a rate $|U_{e4}|^2$ times smaller. Once produced, the $\nu_4$ mass eigenstates immediately decay to a light neutrinos and the scalar boson. The scalar then decays to a neutrino-antineutrino pair, giving rise to our signal. Overall, the process of interest is
\begin{align}
    \nu_4 (E_{\nu_4}) \,\to\, \nu(E_1) \,+\, \phi(E_\phi)&
    \\\nonumber
    &^\searrow \nu (E_2) \,+\,  \overline{\nu} (E_3).
\end{align}
For most cases of interest, $E_{\nu_4} \gg m_4$, so if $\nu_4$ ($\overline{\nu_4}$) is produced via weak interactions, it will be left-handed (right-handed) polarized to a very good approximation. We then assume all heavy neutrinos to be polarized with a definite helicity $h_4=-1$ for neutrinos and $h_4=+1$ for antineutrinos. Nevertheless, due to the assumption of parity conservation for the $\phi$ interactions with neutrinos, both helicity flipping (HF) and helicity conserving (HC) decay channels are allowed. Assuming all neutrinos to be ultra-relativistic, we find the squared-amplitudes for polarized $\nu_4^{h_4=-1} \to \nu^{h} \phi$ decay,
\begin{align}
\left|M_{\nu_4^-\to \nu^-\phi}\right|^2 &= \sum_{i=1}^3 g_\phi^2 \,|U_{si}|^2  m_4^2 \frac{E_1}{E_\nu},
\\
\left|M_{\nu_4^-\to \nu^+\phi}\right|^2 &= \sum_{i=1}^3 g_\phi^2 \,|U_{si}|^2 m_4^2  \left[(1-r_\phi^2)-\frac{E_1}{E_\nu}\right]
\end{align}
where $r_\phi = m_\phi/m_4$. Integrated over phase space, both channels contribute identically to a total decay rate of
\begin{equation}\label{eq:nu4rate}
    \Gamma^{\rm LAB} (\nu_4 \to \nu \phi) = \sum_{i} \frac{g_\phi^2}{16\pi} \frac{m_4^2}{E_\nu} |U_{s4}U_{s i}|^2 (1-r_\phi^2)^2.
\end{equation}
Our decay rate is in agreement with Refs~\cite{Kim:1990km,Dentler:2019dhz}. Note that helicity conserving decays prefer larger $E_1$ values, while helicity flipping decays prefer smaller values of $E_1$. Therefore, for our present application, helicity-flipping decays are important since the antineutrinos from the subsequent scalar decay tend to be more energetic. Also important is the limit $r\to1$, where the scalar particle has most of the $\nu_4$ energy regardless of the helicity structure of the decay. This is the scenario with the most energetic antineutrinos in the final state, for which a simultaneous explanation of MiniBooNE and LSND is most successful.

The scalar decay length in the lab frame to leading order in the small mixing elements is
\begin{equation}
    \Gamma^{\rm LAB} (\phi \to \nu \overline{\nu}) = \sum_{i,j=1}^3\frac{g_\phi^2}{8\pi} \frac{m_\phi^2}{E_\phi} |U_{s i}U_{s j}|^2.
\end{equation}
As expected, the scalar decays are doubly suppressed by small mixing elements, and so it tends to decay more slowly than $\nu_4$. Nevertheless, the decay of both particles can be considered prompt within astrophysical objects. Finally, note that only due to parity conserving nature of the scalar interaction, both left- and right-handed antineutrinos are produced. In this case, only the right-handed antineutrinos ($\overline{\nu}^+$) are relevant for detection through weak interactions.

\section{Solar Antineutrinos}\label{sec:solantinus}

The flux of MeV antineutrinos from the Sun in the SSM is negligibly small. The largest antineutrino flux at MeV energies comes from small fractions of long-lived radioactive isotopes in the Sun, namely $^{232}$Th, $^{238}$U, and mainly $^{40}$K. This give rise to an antineutrino flux on Earth of about $200$~cm$^{-2}$~s$^{-1}$ with $E_\nu \lesssim 3$ MeV~\cite{Malaney:1989hs}. This component, however, is still 6 orders of magnitude smaller than the geoneutrino flux at the surface of the Earth at these energies, and can be safely neglected. At larger energies, photo-fission reactions produce an even smaller flux of antineutrinos of about $10^{-3}$~cm$^{-2}$~s$^{-1}$~\cite{Malaney:1989hs}. It is only down at the much lower energies of tens of keV that antineutrinos start being produced in thermal reactions at a similar rate to neutrinos with fluxes as large as $10^9$~cm$^{-2}$~s$^{-1}$~\cite{Vitagliano:2017odj}.

Existing limits on the flux of solar antineutrinos are usually quoted in terms of an energy-independent probability $P_{\nu_e \to \overline{\nu_e}}$ of conversion of $^8$B neutrinos into antineutrinos. The most stringent limits were obtained by KamLAND in 2011~\cite{Collaboration:2011jza}
\begin{equation}
    P^{\rm KamLAND-2011}_{\nu_e \to \overline{\nu_e}}(E_\nu \geq 8.3\,\, \text{MeV}) < 5.3 \times 10^{-5},
\end{equation}
which was recently improved in 2021~\cite{Abe:2021tkw},
\begin{equation}
    P^{\rm KamLAND-2021}_{\nu_e \to \overline{\nu_e}}(E_\nu \geq 8.3\,\, \text{MeV}) < 3.5 \times 10^{-5},
\end{equation}
and by Borexino in 2019~\cite{Agostini:2019yuq}
\begin{equation}
    P^{\rm Borexino}_{\nu_e \to \overline{\nu_e}} (E_\nu\geq1.8\,\, \text{MeV}) < 7.2\times10^{-5},
\end{equation}
all at $90 \%$ C.L. In addition, SuperKamiokande (SK) has derived limits on extraterrestrial $\overline{\nu_e}$ sources during phases I, II and III~\cite{Bays:2011si}, but the high energy thresholds of $E_\nu > 17.3$ MeV make them irrelevant for the study of $^8$B neutrinos. For SK phase IV (SK-IV), improvements to the trigger system were implemented and the detection of neutron capture on Hydrogen was made possible, lowering thresholds to $E_\nu > 13.3$ MeV~\cite{Zhang:2013tua}. The constraint on solar antineutrino flux was found to be $P^{\rm SK-IV-2013}_{\nu_e \to \overline{\nu}_e} < 4.6\times 10^{-4}$. Recently, further improvements to the neutron tagging algorithm lowered this value to $E_\nu > 8.3$ MeV~\cite{Super-Kamiokande:2020frs}, and using the data 2008 to 2018 the limit was improved to
\begin{equation}
    P^{\rm SK-IV-2020}_{\nu_e \to \overline{\nu_e}} (E_\nu\geq 8.3\,\, \text{MeV}) < 3.6\times10^{-4}.
\end{equation}
A previous preliminary result was shown in Ref.~\cite{Linyan2018} and an even more recent update was presented in Ref.~\cite{alberto_giampaolo_2021_4704606}. Loading of Gd in the SK water tank is expected to greatly improve the neutron tagging efficiency, and would allow for much more stringent limits. With projections on the signal selection efficiency and background reduction, Ref.~\cite{Super-Kamiokande:2020frs} finds that a limit of $P^{\rm SK-IV-Gd}_{\nu_e \to \overline{\nu_e}} (E_\nu\geq 8.3\,\, \text{MeV}) \lesssim 2.2\times10^{-5}$ could be achieved with 0.2\% Gd loading~\cite{Super-Kamiokande:2020frs}. Lowering the energy threshold of the trigger could further improve these projections.

In addition to these, SNO has also set limits at the level of $P^{\rm SNO}_{\nu_e \to \overline{\nu_e}} (E_\nu \in [4,14.8]\,\, \text{MeV}) < 8.3\times10^{-3}$~\cite{Aharmim:2004uf} at $90\%$ C.L. All limits quoted above assume a total $^8$B flux of $5.88 \times 10^{6}$~cm$^{-2}$~s$^{-1}$, except Borexino which assumes $5.46 \times 10^{6}$~cm$^{-2}$~s$^{-1}$, and KamLAND which assumes $5.94 \times 10^{6}$~cm$^{-2}$~s$^{-1}$. At the lowest energies, a bound can also be obtained by noting that the number of elastic $\nu - e $ scattering events in solar neutrino experiments decreases if too many $\nu_4$ states are produced, both due to lower $\overline{\nu_e}-e$ cross sections and suppressed $\nu_e$ flux. These effects, however, are insensitive to variations of the total $\nu_e$ flux below the tens percent level. The predictions from the sterile neutrino decay model are compared with the $^8$B flux in~\reffig{fig:B8flux}. The independent bounds quoted by KamLAND, Borexino, and SK-IV are shown in \reffig{fig:fluxlimits} as a function of $E_\nu$.

The strength of the limits above is mostly due to the large cross section for Inverse beta decay (IBD) on free protons at MeV electron-antineutrino energies. Beyond dominating over the neutrino-electron elastic scattering cross section by about two orders of magnitude (see \reffig{fig:B8flux}), this channel has a distinct signature that drastically reduces backgrounds. After produced, the positron annihilates and the final state neutron is quickly captured by the free protons. This results in a double-bang signal with a positron kinetic energy $T_e \simeq E_\nu - 1.8$ MeV, and a delayed emission of a $\approx 2.2$ MeV gamma. The cross section for this process is well understood at high~\cite{LlewellynSmith:1971uhs} and low~\cite{Vogel:1999zy} energies, and relatively simple formulae that are valid for all energy regimes have been derived by Ref.~\cite{Strumia:2003zx}. In this work we implement the latter calculation, which is provided as machine-friendly data files by Ref.~\cite{Ankowski:2016oyj}.

\begin{figure*}[t]
\centering
\includegraphics[width=0.47\textwidth]{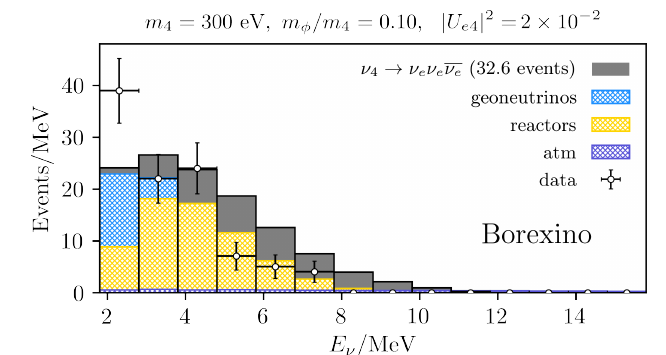}
\includegraphics[width=0.47\textwidth]{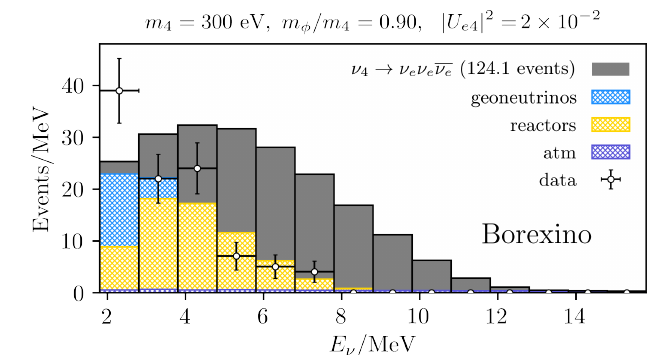} \\
\includegraphics[width=0.47\textwidth]{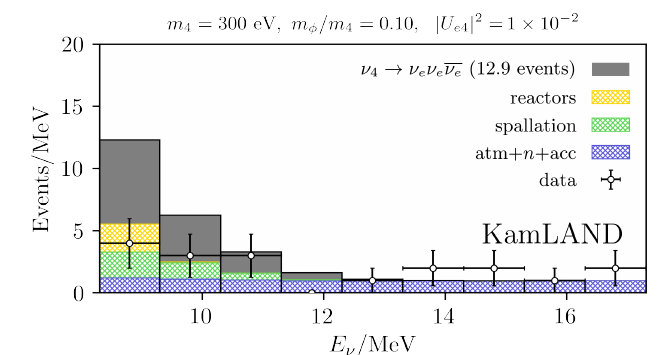}
\includegraphics[width=0.47\textwidth]{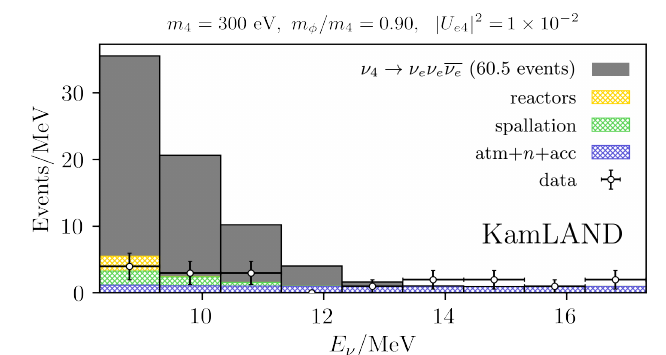}\\
\includegraphics[width=0.47\textwidth]{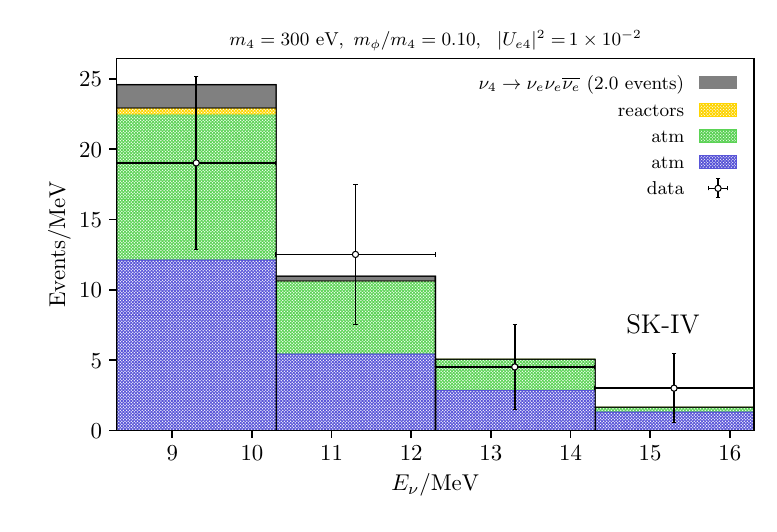}
\includegraphics[width=0.47\textwidth]{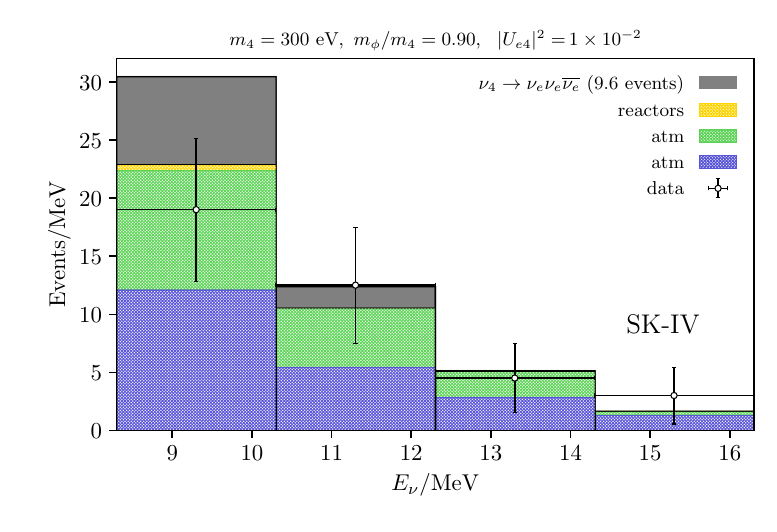} 
\caption{The inverse beta decay spectrum at Borexino (top row), KamLAND 2011 (middle row), and SK-IV 2020 (bottom row), all as stacked histograms. The hatched histograms show the background estimations by the collaboration, and the filled histogram (grey) our prediction of visible neutrino decays. All plots assume $|U_{\mu_4}|=|U_{\tau 4}|=0$. \label{fig:exp_spectra}}
\end{figure*}

\paragraph{Backgrounds} For Borexino, reactor neutrinos represent the largest source of backgrounds, but are effectively constrained by DayaBay measurements. Atmospheric neutrino events with genuine IBD scattering or otherwise inherit large uncertainties from the atmospheric neutrino flux and cross sections, but represent only a small contribution ($6.5\pm 3.2$ events). Finally, the $^{238}$U and $^{232}$Th geoneutrino fluxes are energetic enough to contribute to the IBD sample, but are only significant up to 3.2~MeV. Borexino omits the contribution of geoneutrinos from the Earth's mantle in their estimation, which is conservative. This component is the most likely explanation for the $\approx 2\sigma$ excess seen in the lowest energy bin~\cite{Agostini:2019yuq} (see also their latest geoneutrino analysis~\cite{Agostini:2019dbs}). 

The reactor neutrino flux at KamLAND is dominant below $8.3$ MeV, but contributes only about $2.2$ events above that value. Due to the smaller overburden at KamLAND and SK, they suffer from larger spallation backgrounds, coming mainly from radioactive decays of $^9$Li. After muon tagging and fiducial volume cuts, these are reduced to less than 5 events at both locations. The large number of neutrino-electron scattering events presents a background for SK. For this reason, a cut is applied requiring small shower angles with respect to the direction of the Sun, $\cos{\theta_{\odot}} < 0.9$. This does not impact IBD events as the positron angle with respect to the incoming neutrino is significantly larger ($\langle \cos{\theta}\rangle \approx  0$) than in the predominantly forward process of elastic scattering. The observed event spectra and background predictions by the respective collaborations are shown in \reffig{fig:exp_spectra}.

\subsection{IBD Rates from Decays} 
\begin{figure*}[t]
    \centering
    \includegraphics[width=0.49\textwidth]{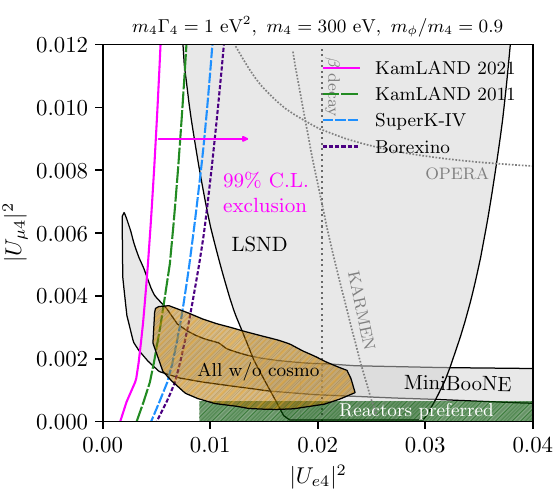}
    \includegraphics[width=0.49\textwidth]{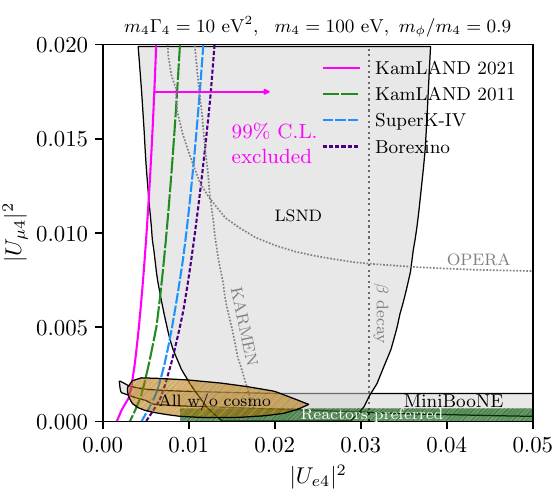}
\caption{\label{fig:limits}Limits from solar antineutrino searches on the active-heavy neutrino mixing at $99\%$ C.L. The regions required to explain the short-baseline anomalies in a Dirac sterile neutrino decay model are also shown ($99\%$ C.L.). For reactors, a preferred interval in $|U_{e4}|^2$ is shown and is independent of $|U_{\mu 4}|^2$. On the left we show the $m_4 \Gamma_4 = 1$ eV$^2$ case and on the right $m_4 \Gamma_4 = 10$ eV$^2$, where the $\nu_4$ is shorter-lived. Our bounds are the same for the two cases.}
\end{figure*}

\begin{figure}[t]
    \centering
    \includegraphics[width=0.49\textwidth]{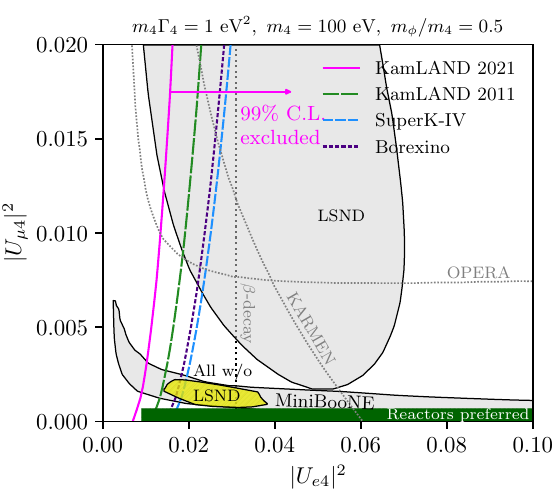}
\caption{\label{fig:limits2}Same as \reffig{fig:limits}, but for $m_\phi/m_4 = 0.5$. In this regime LSND and MiniBooNE are harder to combine due to softer $\overline{\nu}_e$ spectrum obtained via electron mixing. We show only the combination of MiniBooNE with other datasets, excluding LSND, as reported in \refref{Dentler:2019dhz}. The solar antineutrino spectrum is also softer, but can effectively constrain the preferred combined region at $99\%$ C.L.}
\end{figure}

The largest observable flux from sterile neutrinos would come from $\nu_4$ states produced via weak interactions in the decay of $^8$B. The number of IBD events in a given experiment can be computed as
\begin{align}\label{eq:nevents}
 \frac{\dd N_{\rm events}}{\dd E_{\nu_4} \, \dd E_1 \,\dd E_3} = \mathcal{N} 
 \,\frac{\dd \Phi^{\nu_4}}{\dd E_{\nu_4}}\, \frac{\dd P^{\rm dec} }{\dd E_1 \dd E_3}\,\sigma(E_3) \,\left\langle P_{\overline{\hat{\nu_s}}\to \overline{\nu_e}} \right \rangle(E_3),
\end{align}
where $\mathcal{N}$ stands for total exposure of the experiment, $\left\langle P_{\overline{\hat{\nu_s}}\to \overline{\nu_e}} \right \rangle$ is the flavour transition probability for Solar antineutrinos averaged over the radius of the Sun (see \refapp{app:Pse}), and
\begin{align}
\frac{\dd \Phi^{\nu_4}}{\dd E_{\nu_4}} & = |U_{e4}|^2 \frac{\dd \Phi^{\nu_e}}{\dd E_\nu},
\\\nonumber
\frac{\dd P^{\rm dec} }{\dd E_1 \dd E_3} & = \frac{1}{\Gamma_{\nu_4} \Gamma_{\phi}} \frac{\dd \Gamma_{\nu_4\to\hat{\nu_s} \phi}}{\dd E_1} \frac{\dd \Gamma_{\phi\to \hat{\nu_s} \overline{\hat{\nu_s}}}}{\dd E_3}.
\end{align}
Note that \refeq{eq:nevents} is the analogue of Eq.\,(9) from Ref.~\cite{Dentler:2019dhz}, and is simpler since we work with very long baselines and under the assumption that the number of initial $\nu_\mu$ states is negligible. For the flux of $^8$B neutrinos, $\dd \Phi^{\nu_e}/\dd E_\nu$, we implement the high-metallicity fluxes in the SSM AGSS09-B16~\cite{Vinyoles:2016djt}, where the total $^8$B neutrino flux is $5.46 \times 10^{6}$~cm$^{-2}$~s$^{-1}$. For low-metallicity models, our constraints on the new physics coupling are weakened by about $20\%$.

With the predicted number of IBD events at each solar neutrino experiment, we implement our statistical test (described in detail in \refapp{app:statistics}) to place limits on the active-heavy mixing angles. Our $\chi^2$ test statistic models solar neutrino flux and experimental backgrounds uncertainties through bin-uncorrelated nuisance parameters with Gaussian errors. Both flux and background uncertainties are fixed at $10\%$, except for the SK-IV, for which we inflate those to $20\%$. For the KamLAND 2021 dataset, we combine the data into 2-MeV-wide bins when performing our fit. We have also performed a total rate fit for each of the energy bins to obtain model-independent limits on the solar antineutrino flux. We find similar to those provided by the collaborations, within 50\%. Our limits are always weaker, and therefore more conservative, when compared to the official ones. For SK-IV, no model-independent limit was shown in the final article~\cite{Super-Kamiokande:2020frs}, so we show our own result in \reffig{fig:fluxlimits}.

\section{Results}\label{sec:results}

\begin{figure*}[t]
    \centering
    \includegraphics[width=0.49\textwidth]{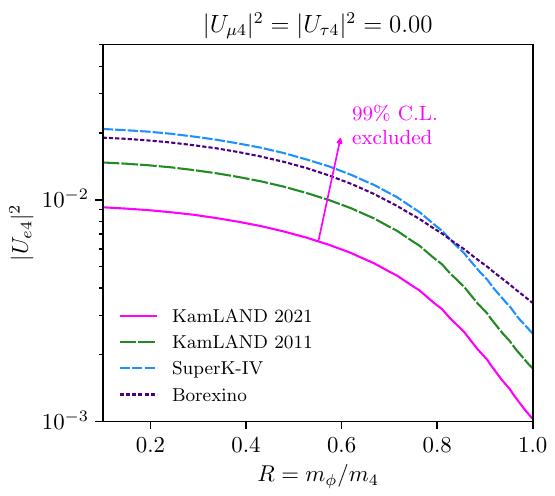}
    \includegraphics[width=0.49\textwidth]{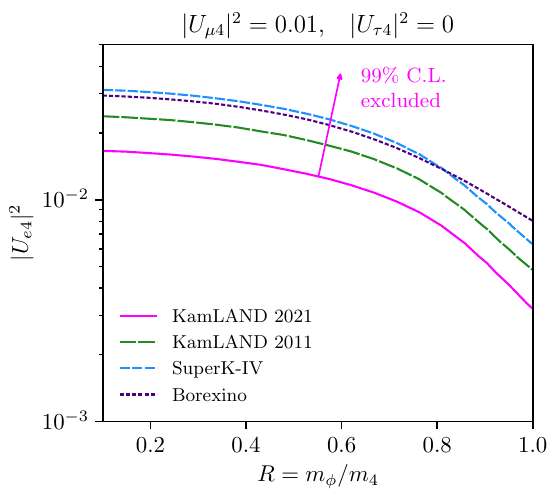}
\caption{\label{fig:limits_ratio} Limits obtained in this work in the plane of active-heavy mixing versus the ratio $R = m_\phi/m_4$, at $99\%$ C.L. On the left we show the case of $m_4 = 300$ eV, but we note that our curves are mostly insensitive to the overall scale of $m_4$ and $m_\phi$ in the region of interest for short-baseline anomalies.}
\end{figure*}

We plot our $90\%$ C.L. limits in \reffig{fig:limits} as a function of $|U_{e4}|^2$ and $|U_{\mu4}|^2$ for $m_\phi/m_4=0.9$. On the same axes, we show the preferred (grey-shaded) regions obtained in Ref.~\cite{Dentler:2019dhz} to explain LSND and MiniBooNE individually, as well as the combined fit to MiniBooNE, LSND and global data (except reactors and cosmology) as ``All w/o cosmo". Weaker constraints from the OPERA~\cite{Agafonova:2013xsk} and KARMEN~\cite{Armbruster:2002mp} neutrino experiments, as well as beta decay kink searches are shown as dashed grey lines.
We pick two particular cases with the shortest $\nu_4$ and $\phi$ lifetimes to compare against our limits, corresponding to $m_4 \Gamma_4= 1$~eV$^2$  and $m_4\Gamma_4=10$~eV$^2$. These lifetimes are achieved for couplings close to the perturbativity limit, namely $g_\phi^2 = (1.5)^2 \times 10^{-2}/({|U_{e4}|^2+|U_{\mu4}|^2)} $ and $g_\phi^2 =  (12)^2 \times 10^{-2}/{(|U_{e4}|^2+|U_{\mu4}|^2)}$, respectively. It is clear that an explanation of LSND is in large tension with solar antineutrino searches for all three experiments we consider. It should also be noted that the region with large $|U_{\mu4}|^2$ which is not excluded by our curves is excluded by MiniBooNE itself. As expected, KamLAND leads to the strongest bounds despite its large neutrino energy thresholds $8.3$ MeV. Borexino and SK-IV bounds are competitive, with the latter performing better for harder antineutrino spectra.

In \reffig{fig:limits2}, we show our constraints for the case $m_\phi/m_4=0.5$. A simultaneous explanation of MiniBooNE and LSND is more challenging now, and only a global fit including MiniBooNE but not LSND is available (``All w/o LSND''). In this case, we constrain the region preferred by MiniBooNE significantly. Lower values of $m_\phi/m_4$ are even more challenging from the point of view of explaining the SBL results, as $\phi$ becomes longer lived and helicity-flipping decays of $\nu_4$, which lead to soft daughter spectra, become more important.

Changing the heavy neutrino mass but keeping the ratio $R = m_\phi/m_4$ fixed leaves our bounds unaltered for all the mass range of interest. Lowering R, on the other hand, weakens our bounds slightly due to the softer solar antineutrino spectrum, although the weakening saturates once below $m_\phi/m_4 \approx 0.5$. We show constraints on the electron mixing angle in \reffig{fig:limits_ratio}. The strongest constraints are obtained for vanishing muon and tau mixing, and for $R\to1$, as in that case $\phi$ carries most of the energy of its parent particle $\nu_4$. These limits are independent of the absolute scale of $m_\phi$ and $m_4$, provided these masses are below the Q-values of the $^8$B decays and above the light neutrino masses.

Accounting for perturbativity bounds on $g_\phi$ and the baselines of LSND and MiniBooNE, a lower bound on $m_4$ of $\approx 100$ eV can be obtained for neutrino mixing of the order of $|U_{e4}|^2\approx10^{-3}$. Below this value, $\nu_4$ is too long-lived and does not lead to interesting signatures at short-baseline experiments. The preferred regions for MiniBooNE and LSND shift to larger mixing angles when either $\nu_4$ or $\phi$ are longer-lived, while our constraints remain unaffected. The parameters used in Figs.~\ref{fig:limits} and \ref{fig:limits2} are chosen so as to minimize the decay length of new particles. Bounds from kink searches in beta decay become the strongest above $m_4 \gtrsim 5$ keV, and peak searches in meson decay preclude an explanation of LSND with the current model at the MeV scale.

We note that a massless scalar has also been discussed as an explanation of the MiniBooNE and LSND anomalies~\cite{deGouvea:2019qre}, although it is disfavored with respect to the best-fit point of Ref.~\cite{Dentler:2019dhz} at $99\%$ C.L. Interpreting that study in the present lepton-number and parity conserving model, one would obtain no constraint from solar antineutrino searches. In that case, however, light neutrinos will also decay. We comment more on this possibility and others in \ref{sec:alternatives}.

\subsection{Future opportunities}

Borexino has collected an additional $\mathcal{O}(500)$ days of data on top of the 2771 days already analyzed in Ref.~\cite{Agostini:2019yuq}. In addition, the improvements made by the collaboration in the latest geoneutrino analysis~\cite{Agostini:2019dbs}, such as enlarged fiducial volume and improved background rejection, can be implemented in the solar antineutrino search. Just in terms of exposure, this represents an improvement of $40\%$ with respect to the values we use. 

\paragraph{Synergy with DSNB} Solar antineutrino searches will become even more stringent with upcoming efforts to detect $\overline{\nu_e}$ events from the DSNB~\cite{Horiuchi:2008jz,Lunardini:2009ya}. The SK detector is expected to detect this neutrino flux with the addition of Gd to its detector volume~\cite{Beacom:2003nk}. The large neutron capture cross sections on Gd and the emission of $8$~MeV gammas will help reduce backgrounds and lower the $\overline{\nu_e}$ detection threshold to neutrino energies as low as the IBD threshold, provided $E_e > 0.8$~MeV~\cite{Simpson:2019xwo}. The increased detection efficiencies at lower energies, and reduced accidental and mis-reconstructed backgrounds will improve on the limits we set, being limited by intrinsic reactor and atmospheric $\overline{\nu}_e$ backgrounds, but also by an exponentially rising spallation background. Future large liquid-scintillator detectors, such as the Jiangmen Underground Neutrino Observatory (JUNO)~\cite{An:2015jdp}
and the Jinping neutrino experiment~\cite{JinpingNeutrinoExperimentgroup:2016nol}, can also improve on current constraints with an expected threshold of $E_\nu \gtrsim 8.5$ MeV~\cite{Li:2019snw}.  In the far future, observatories capable of accumulating larger numbers of DSNB events, such as the proposed detector THEIA~\cite{Askins:2019oqj}, would play an important role in searching for solar antineutrinos. We note that in the event of a detection of the DSNB, one could also constrain the models considered here by requiring small DSNB absorption by relic neutrinos~\cite{Jeong:2018yts,Bustamante:2020mep}. 

\paragraph{Light neutrino decay} Even in the parity conserving model discussed so far, one can avoid solar antineutrinos by resorting to a massless $\phi$. In that case, however, the light mostly-active neutrinos will decay. For typical parameters relevant for the SBL anomalies, this decay will happen within $1$ AU, both visibly and invisibly. For instance, consider $\nu_2 \to \nu_1 \phi$ and $\nu_3 \to \nu_1 \phi$ decays with normal ordering and $m_1\approx0$. For a coupling of $g_\phi=1$, we find
\begin{align}
    c\tau_2^{\rm LAB}&\approx 0.9 \text{ AU} \left(\frac{10^{-5}}{|U_{s1}U_{s2}|^2}\right) \left(\frac{E_2}{10\text{ MeV}}\right),
    \\
    c\tau_3^{\rm LAB} &\approx 0.03 \text{ AU} \left(\frac{10^{-5}}{|U_{s1}U_{s3}|^2}\right) \left(\frac{E_3}{10\text{ MeV}}\right). \label{eq:faster_kind}
\end{align}
In the convention adopted by the literature, these correspond to $\tau_2^{\rm 0}/m_2 = 4.4\times10^{-5}$~s/eV and $\tau_3^{\rm 0}/m_3 = 1.3\times10^{-6}$~s/eV. For inverted ordering, both $\nu_1\to\nu_3\phi$ and $\nu_2\to\nu_3\phi$ decays are of the faster kind in \refeq{eq:faster_kind}. On top of the cosmological issues with such short lifetimes (for recent discussions, see Refs.~\cite{Escudero:2019gfk,Escudero:2020ped}), the largest $g_\phi$ values relevant for the allowed regions are already excluded by laboratory experiments, such as SNO~\cite{Aharmim:2011vm}, which is consistent with no disappearance of solar neutrinos (see Refs.~\cite{Beacom:2002cb,Berryman:2014qha}). Other datasets have also been discussed to constrain the lifetime of light neutrinos, including measurements of the flavor ratios of cosmic neutrinos~\cite{Bustamante:2016ciw} and of the Glashow resonance~\cite{Bustamante:2020niz}. It should be noted, however, that light-sterile mixing parameters governing light neutrino decay are related to those of SBL anomalies in a model-dependent fashion. In principle, but not without fine-tuning, the correlation between $|U_{e4}|$, $|U_{\mu4}|$, $|U_{s4}^*U_{sj}|$, and $|U_{si}^*U_{sj}|$ for $i,j<4$ may be relaxed. 
In models with LN violation or LN charged scalars (see below), provided several constraints are satisfied, solar antineutrinos may become relevant again for massless $\phi$ as the light-neutrino decays $\nu_2\to \overline{\nu_1} \phi$ are open.

\section{Alternative models: violating parity and lepton number}
\label{sec:alternatives}
Various other possibilities for visible sterile neutrino decay exist, depending on the Dirac or Majorana nature of neutrinos, as well as on the parity structure of the sterile neutrino sector. While we only focused on the parity conserving model discussed above, we would like to dedicate this section to understanding if other minimal extensions of the SM by a singlet sterile neutrino and a scalar are subject to our constraints. For clarity, we focus on SM extensions with a single new sterile neutrino: $\nu_s=\nu_s^L + \nu_s^R$ in the Dirac case and $\nu^R$ in the Majorana case. Our findings for the minimal models are summarized in \reftab{tab:minimal_models}.

\renewcommand{\arraystretch}{1.8}
\begin{table*}[t]
    \centering
    \begin{tabular}{|c|c|c|c|c|c|}
    \hline
        Minimal Models & Parametric Limit & Polarized $\nu_4^-$ decay & Scalar decays & Spectrum of ${\color{purple} \overline{\nu}^{+}}$ & Expected Signals \\\hline\hline
       \multirow{3}{*}{ \shortstack{Dirac $L(\phi)=0$\\$\mathscr{L}_{\rm int}^{D-0}$}}%
       & $|(V_L)_{s4}(V_R)_{si}|\sim
|(V_R)_{s4}(V_L)_{si}|$ & $\nu_4^- \to \nu^{-} \phi$ / $\nu^{+}\phi$ & $\phi\to \nu^-\overline{\nu}^-$ / $\nu^+{\color{purple} \overline{\nu}^{+}}$ & Hard & Solar/SBL  \\
        &$|(V_R)_{s4}(V_L)_{si}|\to0$ & $\nu_4^-\to \nu^{+}\phi$ & $\phi\to \nu^+{\color{purple} \overline{\nu}^{+}}$ & Soft & Solar/SBL \\
        &$|(V_L)_{s4}(V_R)_{si}|\to0$  & $\nu_4^-\to \nu^{-}\phi$ & $\phi\to \nu^-\overline{\nu}^-$ & None & SBL \\
\hline\hline
       \multirow{3}{*}{ \shortstack{Dirac $L(\phi)=-2$ \\ $\mathscr{L}_{\rm int}^{D-2}$}}%
       & $g_L \sim g_R$ & $\nu_4^-\to \overline{\nu}^{-}\phi^*$ / ${\color{purple} \overline{\nu}^{+}}\phi^*$ & $\phi^*\to \nu^-\nu^-$ / $\nu^+\nu^+$ & Hard & Solar/SBL \\
        &$g_R \to 0$ & $\nu_4^-\to{\color{purple} \overline{\nu}^{+}}\phi^*$ & $\phi^*\to \nu^-\nu^-$ & Hard & Solar/SBL \\
        &$g_L \to 0$ & $\nu_4^-\to \overline{\nu}^{-}\phi^*$ & $\phi^*\to \nu^+\nu^+$ & None & None \\
\hline\hline
        \multirow{2}{*}{\shortstack{Majorana \\ $\mathscr{L}_{\rm int}^M$ 
        }}
        & \multirow{2}{*}{$-$} & \multirow{2}{*}{$\nu_4^-\to \nu^{-} \phi$ / ${\color{purple} \overline{\nu}^{+}}\phi$} & \multirow{2}{*}{$\phi\to \nu^-\nu^-$ / ${\color{purple} \overline{\nu}^{+}}{\color{purple} \overline{\nu}^{+}}$} & \multirow{2}{*}{Hard} & \multirow{2}{*}{Solar/SBL} \\
        & & & & & \\
        \hline
        \end{tabular}
    \caption{Minimal models for sterile neutrino decay to light neutrinos. Here $\nu$ and $i$ stand for all light neutrinos mass eigenstates ($i<4$), and $\nu^-_{4}$ for the heavier left-handed polarized neutrino. The first two columns show the minimal model considered and the judicious choices of its parameters to achieve a certain parity structure. The third and fourth columns show the decay channels allowed in that model, separated by all possible helicity final states. The penultimate column shows the kind of visible solar antineutrinos (${\color{purple} \overline{\nu}^{+}}$) energy spectrum is predicted in the model. This depends on the HF or HC nature of the $\nu_4^-$ decay. Note, however, that the spectrum may change in the limit $m_\phi/m_4\to 1$. For the minimal Majorana neutrino model, one always obtains a prediction for visible solar antineutrinos. See the main text for definitions.}
    \label{tab:minimal_models}
\end{table*} 

\paragraph{Dirac $L(\phi)=0$} We start with a generalization of \refeq{eq:Lscalar} by writing
\begin{align}\label{eq:LD0}
     \mathscr{L}_{\text{int}}^{D-0} &= g_\phi \overline{\nu^L_s}\nu^R_s\phi + \text{h.c.} 
     \\\nonumber
     &= g_\phi (V_L)_{si}^*(V_R)_{sj} \overline{\nu_i}P_R\nu_j \phi + \text{h.c.},
\end{align}
where index summation is understood. For complex $g_\phi$, this is the most generic parametrization of the vertex. We implicitly diagonalized the Dirac mass matrix by means of two unitary matrices $V_L$ and $V_R$, defined by $\nu^L_{s} = (V_{L})_{si} \nu^L_i$, $\nu^R_s = (V_{R})_{si} \nu_i^R$, where $\nu_i^{L,R}$ are the (chiral) mass eigenstates. Note that the enlarged PMNS matrix is defined as $U_{\rm PMNS^\prime} = V_L$ when the charged lepton mass matrix is diagonal. Abandoning the assumption of parity conservation in the sterile sector that was made previously, $V_L=V_R$, one can have allow for $V_L \neq V_R$ by choosing different Yukawa couplings for $\nu_s^L$ and $\nu_s^R$. By breaking parity at the level of the Dirac mass matrix, it is possible to independently tune the couplings appearing in the operators $\overline{\nu_i} P_R \nu_4$ and $\overline{\nu_i}P_L\nu_4$. In practice, this allows to tune the rate for visible and invisible decays of $\nu_4^-$ neutrinos. The same is true for the decay of the scalars, which can be either visible or invisible, depending on $V_L$ and $V_R$. In these models, a connection to the SBL anomalies through visible decays always predicts visible solar antineutrinos provided $\phi$ is heavy enough to decay.

\paragraph{Dirac $|L(\phi)|=2$} One can also introduce scalars carrying LN. These type of scalars have been usually discussed in the context of Majoron models, but for our current purposes, we assume no particular connection to neutrino masses. We consider a model with a Dirac field $\nu_s$, and a complex scalar $\phi$ carrying lepton number $L(\phi)=-2$. In all generality, we can write
\begin{align}\label{eq:LD2}
     \mathscr{L}_{\text{int}}^{D-2} &= g_L \overline{(\nu_s)^c} P_L \nu_s \phi + g_R \overline{(\nu_s)^c} P_R \nu_s \phi + \text{h.c.} 
     \\\nonumber
     &= g_L (V_L)_{si}(V_L)_{sj} \overline{\nu_i}P_L\nu_j \phi 
     \\\nonumber &\qquad + g_R (V_R)_{si} (V_R)_{sj} \overline{\nu_i}P_R\nu_j \phi +\text{h.c.},
\end{align}
where again we implicitly diagonalized the Dirac mass matrix with $V_L$ and $V_R$. In this case, even for parity conserving mass matrices, one can violate parity in the neutrino-$\phi$ interactions by tuning the arbitrary $g_L$ and $g_R$ couplings. In this case, 
$\nu_4^-$ states produced in the Sun will always decay to visible antineutrinos provided $g_R\neq0$, independently of the decay products of $\phi^*$. For this model, one may attempt to explain SBL anomalies with only the decay products of scalar produced in $\nu_4^-$ decays by setting $g_R \to 0$. In that case, no visible solar antineutrinos appear. 

\paragraph{Majorana neutrinos} A final possibility is to abandon LN and work with Majorana neutrinos. In this case, a minimal model can be built with only $\nu^R$ and a scalar $\phi$. LN is violated by the $\nu^R$ Majorana mass term, and the most general interaction Lagrangian in this case is 
\begin{align}\label{eq:Lmaj}
     \mathscr{L}_{\text{int}}^{\rm M} &= g_R \overline{(\nu^R)^c} \nu^R \phi + \text{h.c.} 
     \\\nonumber
     &= g_R (V)_{si} (V)_{sj} \overline{\nu_i}P_R\nu_j \phi +\text{h.c.},
\end{align}
where now we implicitly diagonalized the Majorana mass matrix by means of a single unitary matrix $V=U_{\rm PMNS^\prime}$. In this case, all light neutrinos as well as antineutrinos are visible due to the reduced number of degrees of freedom. Both HC and HF decays of $\nu_4$ are controlled by the same parameters, and cannot be disentangled as easily. Solar antineutrinos could appear in this case if all other constraints are satisfied.

\paragraph{Simplified models} Finally, we note that Refs.~\cite{PalomaresRuiz:2005vf,deGouvea:2019qre} work with simplified models, and do not specify the origin of the $ \mathcal{L}\supset g_e \overline{\nu_e} \nu_4 \phi$ vertex. Although this operator may arise from a Lagrangian as simple as \refeq{eq:Lscalar}, it can be considered more generically as a by-product of non-renormalizable operators such as $(LH)^2 \phi$ and $(LH)\nu_s\phi$. In these effective models, the active-heavy mixing necessary for $\nu_4$ production in most accelerator experiments, $|U_{\mu4}|^2$, is independent from $g_e$, which controls the decay rate of $\nu_4\to\nu_e\phi$. In this case, the only way to generate $\overline{\nu_e}$ appearance at LSND is via muon decays, $\mu^+\to e^+\nu_e\overline{\nu_4}$. It also follows that the mixing $|U_{e4}|^2$ may be parametrically small, turning off $\nu_4$ production in the Sun via mixing. Four-body decays of the type $^8\text{B}\to {^8}\text{Be} \,e^+ \nu_4 \phi$ are negligible as kaon decays constrain $g_e^2 |U_{\mu 4}|^2 < \mathcal{O}(10^{-7})$.

If a vector particle is introduced instead, the cosmological history is yet even more involved. We do not study this case here, although our solar antineutrino bounds would also apply to parity-conserving scenarios with small modifications. Note that our constraints are not relevant for fully invisible sterile neutrino decays, as invoked to relax the tension between SBL appearance and disappearance tension in Refs~\cite{Diaz:2019fwt,Moulai:2019gpi}.

\section{Conclusions}\label{sec:conclusions}

Puzzling results from some of the short-baseline neutrino experiments will eventually find an explanation with more data coming from the SBN program at Fermilab~\cite{Cianci:2017okw,Antonello:2015lea,Machado:2019oxb} and the $\pi^+$ decay-at-rest experiment at J-PARC, JSNS$^2$~\cite{Ajimura:2017fld}. At the moment it is possible to speculate that some form of new physics in the neutrino sector is responsible for the deviation of LSND and MiniBooNE results from theoretical expectations within the minimal three-generation neutrino model. Among such speculations is the class of models where the excess of antineutrinos at LSND and excess of low-energy electron-like events at MiniBooNE is due to a prompt production and decay of dark sector particles. This new sector is likely to comprise a heavier, mostly sterile neutrino, that can be produced via neutrino mixing in meson decays and nuclear reactions. Such heavier neutrino can generate a cascade decay to an unstable bosonic mediator and light neutrino, giving rise to the admixture of electron antineutrinos in the flux at the end of the decay chain. 

We have shown that up to some model dependence one should expect that regular nuclear processes in the Sun create an antineutrino flux. Such flux is stringently constrained by most of the solar neutrino experiments, at a $O({\rm few}\times 10^{-5})$ level owing to a larger cross sections for the IBD processes, and additional structure to the signal that has been exploited to cut on backgrounds. After application of these constraints, our results disfavor large part of the parameter space of the model in Ref. \cite{Dentler:2019dhz}, and disfavor this mechanism as an explanation of the LSND excess, while significantly narrowing possible parameter space for the MiniBooNE excess. The simulations used to produce the results in this paper are publicly available on \textsc{g}it\textsc{h}ub\footnote{~\href{https://github.com/mhostert/SolarDecayingSteriles}{{\color{blue-violet}\faGithub}\, github.com/mhostert/SolarDecayingSteriles}.}.

In general, our limits disfavor large $\nu\to\overline{\nu}$ transitions that could improve the combined fit of LSND and MiniBooNE data. Such transitions could in principle be avoided if $\phi$ is lighter than the lightest neutrino state, in which case, mixing angles and CP phases have to be fine-tuned to avoid light neutrino decays. The alternative models with parity violation or apparent LN violation presented in \refsec{sec:alternatives} may avoid $\nu\to\overline{\nu}$ transitions even for massive $\phi$, but would require a case-by-case study of the SBL physics and additional constraints. 

Our constraints add to the existing list of problems of the decaying sterile neutrino solutions to the SBL puzzle. Chiefly among them is cosmology and astrophysics. As is well known, new and relatively strongly interacting states can be populated by the thermal processes leading to the modifications of observed quantities, such as primordial nucleosynthesis yields and/or total amount of energy density carried by neutrinos at late times. In addition, these models are likely to cause strong modifications to the supernovae neutrino spectrum. One reason for such modification is the possibility of the neutrino number-changing processes, such as $\nu\nu\to \nu\nu\varphi\to \nu\nu\nu\overline{\nu}$. Given relatively strong couplings in the models of Refs.~\cite{deGouvea:2019qre,Dentler:2019dhz}, the underlying cross sections are far greater than weak interaction cross section, meaning that the neutrinos can share energy and maintain their chemical equilibrium immediately after they leave the star. The main physical effect, the degrading of average energy for the SN neutrinos, can be constrained with the observed signal of SN1987A. This has been explored to constrain neutrino self-interactions inside supernovae by the requirement that neutrinos carry sufficient energy to the outer layers of the collapsing star~\cite{Shalgar:2019rqe}. We point out, however, that a more general statement can be made regarding neutrino energy loss outside the dense environment, which is independent of the explosion mechanism. Details of this effect will be addressed in a future publication.

\begin{acknowledgments}
The authors would like to thank Linyan Wan and Sandra Zavatarelli for correspondence on the SK-IV and Borexino experimental capabilities. We also thank Ivan Esteban and Joachim Kopp for discussions. MP is supported in part by U.S. Department of Energy (Grant No. desc0011842). This research was supported in part by Perimeter Institute for Theoretical Physics. Research at Perimeter Institute is supported by the Government of Canada through the Department of Innovation, Science and Economic Development and by the Province of Ontario through the Ministry of Research, Innovation and Science.
\end{acknowledgments}

\appendix

\section{Solar Flavor Transitions}\label{app:Pse}
\begin{figure*}[t]
    \centering
    \includegraphics[width=0.32\textwidth]{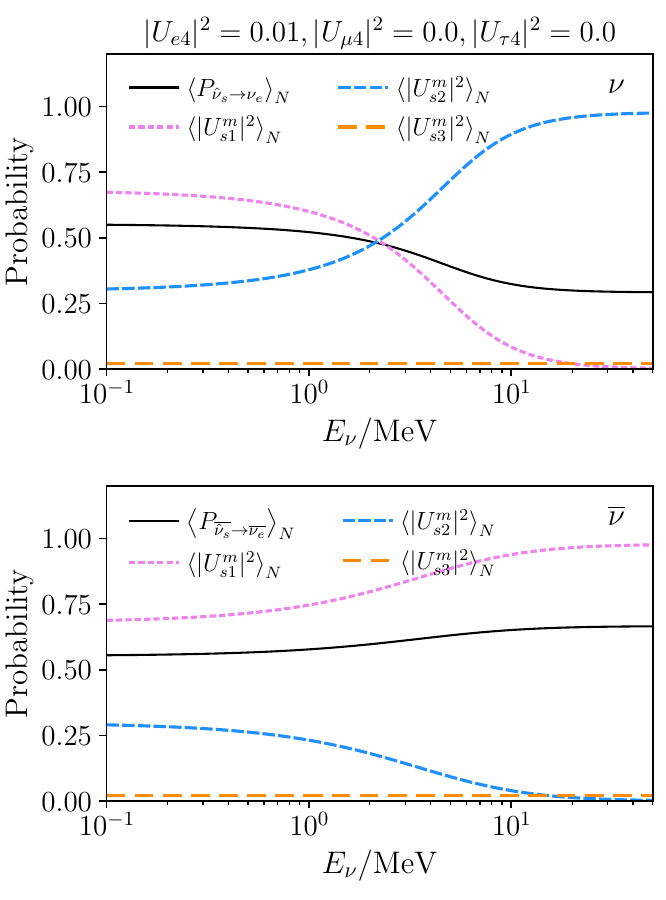}
    \includegraphics[width=0.32\textwidth]{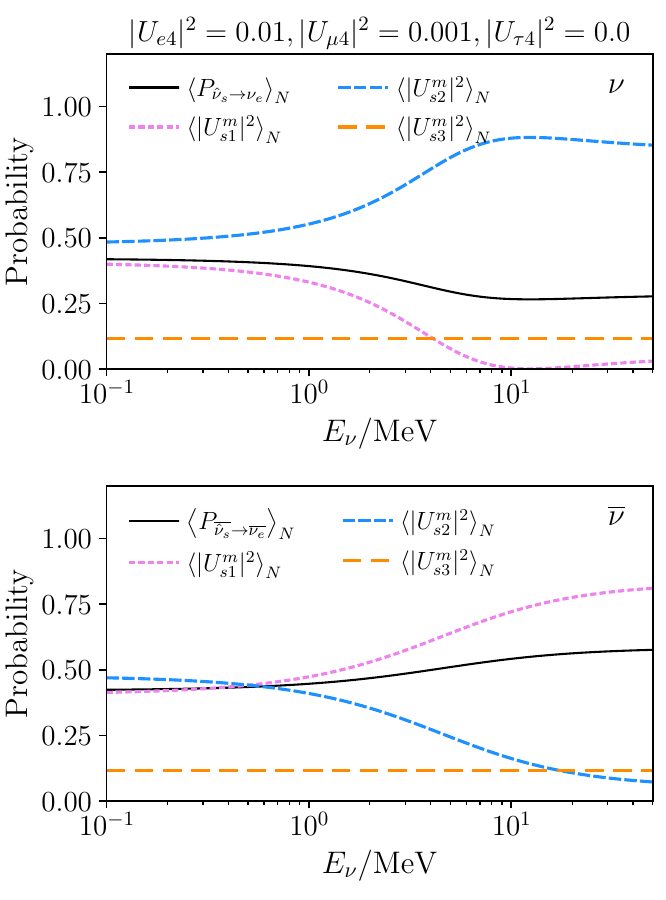}
    \includegraphics[width=0.32\textwidth]{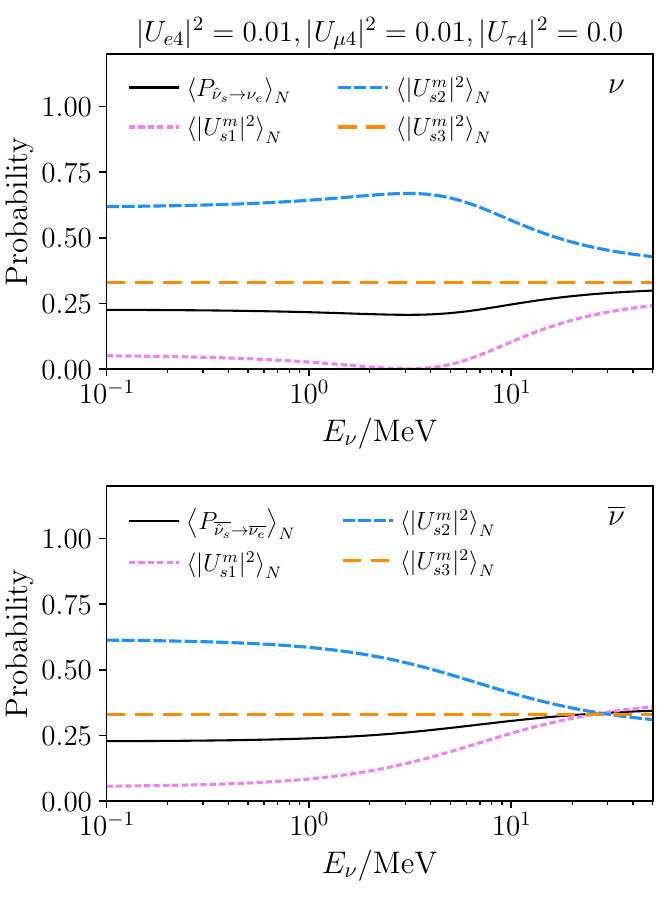}
    \caption{Full flavor transition probability as well as the relevant mixing matrix element in matter, all averaged over the $^8$B neutrino production region in the Sun and properly normalized. From left to right, we increase the muon-heavy mixing, showing neutrino transitions on the top row and antineutrino transitions on the bottom.}
    \label{fig:probability}
\end{figure*}

When $\phi$ decays to light neutrinos, it decays into the state $\ket{\hat{\nu}_s} = \sum_{i=1}^3 U_{s i}^* \ket{\nu_i}$. The average transition probability for $\hat{\nu}_s$ to exit the Sun and be detected as a $\nu_e$ on the surface of the Earth under the (good) approximation of adiabatic flavor conversion is simply 
\begin{equation}\label{eq:pse}
    \langle P_{\hat{\nu}_S\to \nu_e} \rangle_N = \frac{ \left\langle\sum_{i=1}^3 |U_{si}^m|^2|U_{ei}|^2\right\rangle}{|U_{e 4}|^2+|U_{\mu 4}|^2+|U_{\tau 4}|^2},
\end{equation}
which depends on the mixing matrix elements in matter $|U_{si}^m|^2$ at the production point. Here, $\langle \dots \rangle$ denotes a weighted average over the production region and the subscript $N$ refers to taking the non-canonical normalization of $\hat{\nu}_s$ into account. We have neglected Earth matter effects, which for antineutrinos leads to a reduction (increase) of $P_{2\to e}$ ($P_{1\to e}$) below $\sim10\%$, and assumed the unitarity of the $4\times4$ mixing matrix $U$. Neglecting all CP phases, we follow Ref.~\cite{Palazzo:2011rj} and write $U = R_{23} R_{34}R_{24}R_{14} R_{13} R_{12}$, where $R_{ij}=R(\theta_{ij})$ is the usual rotation matrix in the $(i,j)$ plane. Note that if we assume $U_{\tau 4}=-c_{34}s_{23}s_{24}+c_{23}s_{34}=0$, then to leading order in the small angles, $|U_{e4}|^2\sim \theta_{14}^2$ and $|U_{\mu4}|^2\sim c_{23}^2\theta_{24}^2$, with the rest of the mixing matrix elements of the active $3\times 3$ sub-matrix retaining their usual definition.

In the limit where $\Delta m^2_{21}\ll\Delta m^2_{31}\ll\Delta m^2_{41}$, the flavor evolution of solar neutrinos can be described by an effective two-neutrino model, with in-matter modifications only on $\theta_{12}$. Up to corrections proportional to the new mixing angles ($\theta_{14}$, $\theta_{24}$, and $\theta_{34}$) as well as to $\theta_{13}$, the new mixing angle in matter is
\begin{equation}
    \tan{2\theta^M_{12}} \simeq \frac{\sin{2 \theta_{12}} \Delta m^2_{12} }{\Delta m^2_{12} \cos{2 \theta_{12}} - A_{\rm CC}},
\end{equation}
with $A_{\rm CC} = \pm 2\sqrt{2} E_\nu G_F N_e(r)$ for neutrinos (antineutrinos) proportional to the electron density at the production region. Since the sterile component does not feel neutral-current interactions, the neutral-current potential also modifies the relation above. Terms proportional to $A_{\rm NC} = \mp \sqrt{2} E_\nu G_F N_{n}(r)$, where $N_n(r)$ is the neutron number density in the Sun, are small, however. This is because they are proportional to the new small mixing angles, as well as due to the smaller number of neutrons in the Sun, $N_{\rm n}/N_{e} \lesssim 1/2$. In our simulation, we follow Ref.~\cite{Palazzo:2011rj} and keep all corrections in $\theta_{14}$, $\theta_{24}$, $\theta_{34}$, and $\theta_{13}$ in the transition probabilities. As an approximation, we assume the neutral-current potential to follow the same radial dependence as the charged-current one, with $A_{\rm NC}/A_{\rm CC} = -1/4$.

The full transition probabilites using \refeq{eq:pse} in the energy region of the $^8$B flux are shown in Fig.~\ref{fig:probability} for both neutrinos and antineutrinos. In the limit $|U_{\mu4}|,|U_{\tau4}|\to0$, the scalar decays produce mostly-$\nu_e$ states, and $P_{\hat{\nu}_s\to\nu_e}\simeq P_{\nu_e\to \nu_e}$. In this case, the flavor evolution is similar to the standard MSW effect, where neutrino (antineutrinos) undergo resonant (non-resonant) adiabatic flavor conversion, in which production of $\nu_2^m$ at the center of the Sun is enhanced (suppressed). For $|U_{\mu4}|,|U_{\tau4}|\neq0$, the situation is more complex, but the high energy behavior can be understood by taking the limit $\theta_{12}\to \pi/2$ $(0)$ in the mixing elements $|U_{si}|^2$ for neutrinos (antineutrinos). Note that $\nu_3^m \sim \nu_3$, as it should be since $\Delta m^2_{21} \ll \Delta m^2_{31}$.

\section{Statistical Method}\label{app:statistics}

When deriving upper limits on the mixing angles, we minimize the following log-likelihood function  
\begin{align}
\mathcal{L} = 2 &\sum_{i} \left[ \mu_i (\vec{\theta},\vec{\beta}) - D_i + D_i \ln{\frac{D_i}{\mu_i(\vec{\theta},\vec{\beta})}} \right]
\\\nonumber
&+ \sum_{i,j} \frac{\beta_{i,j}^2}{\sigma_{i,j}^2},
\end{align}
where $\vec{\theta}$ stands for the vector of physics parameters (\eg, $|U_{\alpha}|^2$), $\vec{\beta}$ the vector of nuisance parameters with individual entries $\beta_j$ and associated Gaussian errors $\sigma_j$. As an approximation, we assume $\mathcal{L}$ to follow a $\chi^2$ distribution when estimating our confidence intervals. 

The most important systematics for our study are the uncertainties on the total $^8$B solar neutrino flux and total backgrounds numbers. To be conservative, we assign each energy bin two normalisation systematics, one exclusive to the new physics prediction, modelling uncertainties in the solar flux, and one exclusive to backgrounds. All normalization systematics are assumed to be uncorrelated, which is conservative, and are assigned $10\%$ Gaussian errors.

\section{Polarized decay rates}

To produce \reftab{tab:minimal_models}, we computed the decay rates in each channel explicitly. We collect all results for $\nu_4$ and $\phi$ decays assuming massless neutrino final states in each one of the models discussed. The total decay rate for $\nu_i^{h_i}\to\overset{\scriptscriptstyle(\text{\textemdash})}{\nu_j}^{h_j} \phi$ can be obtained by summing each polarized matrix element as
\begin{align}
    \Gamma_4 &= \int_{x^{\rm min}}^{x^{\rm max}} \frac{\sum_{h_i,h_j}|M_{h_i h_j}|^2}{m_4 \beta 16 \pi }\dd x,
\end{align}
where $\beta=|p_{4}|/E_{4}$ is the velocity of $\nu_4$ in the laboratory frame, and $x^{\rm max}_{\rm min} =  (1\pm\beta)(1-r_\phi^2)/2$. Similarly, for $\phi\to\overset{\scriptscriptstyle(\text{\textemdash})}{\nu_i}^{h_i}\overset{\scriptscriptstyle(\text{\textemdash})}{\nu_j}^{h_j}$ decays,
\begin{align}    
\Gamma_\phi &= \int_{x_\phi^{\rm min}}^{x_\phi^{\rm max}} \frac{\sum_{h_i,h_j}|M_{h_i h_j}|^2}{ m_\phi 16\pi} \dd x_\phi,
\end{align}    
where $\beta_\phi=|p_\phi|/E_\phi$ is the $\phi$ velocity in the laboratory frame and $(x_\phi)^{\rm max}_{\rm min} = (1\pm \beta_\phi)/2$.

\subsection{Dirac $L(\phi)=0$ case}

Making use of \refeq{eq:LD0} and neglecting light neutrino masses, the amplitude squared for $\nu_i^{h_i}\to\nu_j^{h_j} \phi$ decays is given by
\begin{align}
    |M_{--}|^2 &= |g_\phi\,(V_L)_{si} (V_R)_{sj}|^2 m_i^2 C(x)
    \\
    |M_{++}|^2 &= |g_\phi\,(V_R)_{si} (V_L)_{sj}|^2 m_i^2 C(x),
    \\
    |M_{-+}|^2 &= |g_\phi\,(V_R)_{si} (V_L)_{sj}|^2 m_i^2 F(x),
    \\
    |M_{+-}|^2 &= |g_\phi\,(V_L)_{si} (V_R)_{sj}|^2 m_i^2 F(x),
\end{align}
which are also valid for $\overline{\nu}_i^{h_i}\to\overline{\nu_j}^{h_j} \phi$ decays. We have defined
\begin{align}
    C(x) &=\frac{2 x - (1-\beta)(1-r_\phi^2)}{2\beta},\\
    F(x) &=\frac{(1+\beta)(1-r_\phi^2)-2x}{2\beta},
\end{align}
which apply for helicity conserving and helicity flipping channels, respectively. Note that $C(x)$ grows while $F(x)$ decreases monotonically with $x$.
For scalar decays $\phi \to \nu^{h_i}_i \overline{\nu}^{h_j}_j$, we compute $|M_{h_1 h_2}|^2$ to find
\begin{align}
    |M_{--}|^2 &= |g_\phi\,(V_L)_{si} (V_R)_{sj}|^2 m_\phi^2,
    \\
    |M_{++}|^2 &= |g_\phi\,(V_R)_{si} (V_L)_{sj}|^2 m_\phi^2,
\end{align}
with all other combinations vanishing in the limit of massless final states. 

\subsection{Dirac $|L(\phi)|=2$ case}

Now, switching to \refeq{eq:LD2}, the amplitudes for $\nu_i^{h_1}\to\overline{\nu}_j^{h_2} \phi$ decays are
\begin{align}
    |M_{--} |^2 &= 4|g_R \,(V_R)_{si} (V_R)_{sj}|^2 m_i^2 C(x),
    \\
    |M_{++}|^2 &= 4|g_L \,(V_L)_{si} (V_L)_{sj}|^2 m_i^2 C(x),
    \\
    |M_{-+} |^2 &= 4|g_L \,(V_L)_{si} (V_L)_{sj}|^2 m_i^2 F(x),
    \\
    |M_{+-}|^2 &= 4|g_R \,(V_R)_{si} (V_R)_{sj}|^2 m_i^2 F(x).
\end{align}
The amplitudes for $\overline{\nu}_i^{h_1}\to\nu_j^{h_2} \phi$ decays can be obtained with the substitution $L\leftrightarrow R$. For scalar decays $\phi \to \nu^{h_1}_i \nu^{h_2}_j$, we compute $|M_{h_1 h_2}|^2$ to find
\begin{align}
    |M_{--}|^2 &= 4 |g_L\,(V_L)_{si} (V_L)_{sj}|^2 m_\phi^2,
    \\
    |M_{++}|^2 &= 4 |g_R\,(V_R)_{si} (V_R)_{sj}|^2 m_\phi^2,
\end{align}
where again the amplitudes for decays into antineutrinos can be obtained by $L\leftrightarrow R$.

\subsection{Majorana case}

In the Majorana case, one can make use of the expressions for the Dirac $L(\phi)=0$ case, keeping in mind that $V_R=V_L$, and that an additional overall factor of $2$ should be included for the $\nu_i$ total decay rates and an overall factor of $2/(1+\delta_{ij})$ for the total $\phi\to \nu_i \overline{\nu_j}$ decay rate.
\vfill
\bibliographystyle{apsrev4-1}
\bibliography{main}{}

\end{document}